# Electro-optic modulator based on Vanadium dioxide epsilon-near-zero vertical nanowire triple-cavity in Silicon Waveguide.

*Gregory Tanyi, Christina Lim and Ranjith Unnithan*

Gregory Tanyi, Christina Lim and Ranjith Unnithan
Department of Electrical and Electronic Engineering, Faculty of Engineering and Information Technology, The University of Melbourne.Parkville VIC 3010, Australia.
 Email: gtanyi@student.unimelb.edu.au, aussunnysun@gmail.com, r.ranjith@unimelb.edu.au, chrislim@unimelb.edu.au.

We present an electro-optic modulator exploiting an array of Vanadium dioxide nanorods operating in the epsilon near regime as the active switching material integrated in a silicon waveguide. The modulator takes advantage of the insulator-to-metal transition of Vanadium dioxide to achieve a robust modulation depth of 19.7 dB/$\square$m with a broad wavelength of operation. Using simulations, we demonstrate how the effective permittivity of the metamaterial can be tuned to a near-zero value by varying the nanorod geometry. The paper also proposes a novel hexagonal array design which achieves low insertion losses while retaining a strong modulation depth. The results provide insight into the design of ultra-compact modulators with high operation frequencies and low insertion losses.

# INTRODUCTION

The huge spike in the demand for computing power has led to a bottleneck in network interconnects. As such, silicon photonic devices have gained significant interest in the past 3 decades given their CMOS compatibility, large bandwidth, high transmission speeds as well as low insertion losses. Optical modulators are key components of photonic integrated circuits which drive optical interconnects. There is a demand for novel modulators because of limitations of current modulators. Silicon, as a material though, is not ideally suited for light emission or modulation because of its weak electrooptic properties which lead to very bulky devices with a high capacitance. Si modulators combined with ring resonators with high Q-factors tend to achieve stronger modulation but are very sensitive to temperature variations and operate over narrow bandwidths. Plasmonic modulators provide strong electro-optic effects because of their strong field confinement but are inherently lossy because of the metals involved.

Mott insulators are a special class of materials which can show unity order index changes across the bulk material under certain conditions. $VO_2$ is a canonical Mott material which exhibits a first order insulator-metal transition (IMT) under certain conditions along with a large change in refractive index as it switches from its insulator phase (2.45+0.509i) to its metallic phase (n=2.04+2.9i). This phase transition can be triggered via electrical, optical and thermal excitations. For our modulator, we focus on the electrical excitation with a field strength of 6.5×107 V/m required to drive the phase change of $VO_2$. The phase transition occurs on a femtosecond scale (28fs) making VO2 an excellent choice as a modulating material in high-speed modulators [9,14].

Recently, there has been a huge spike in the design of modulators based on plasmonic metamaterials due to strong exhibition of nonlinear effects as well as their high speed of operation [6-8]. These nonlinear effects are particularly enhanced in the epsilon-near-zero regime where a component of the real part of the effective permittivity tensor of the metamaterial is near zero [4,7]. The operation of the metamaterial in this regime is dependent on its geometry and can be tuned to achieve superior performance at a chosen wavelength. To circumvent these constraints, plasmonic- based devices (surface plasmon polaritons (SPPs)) have been integrated in silicon photonics [16,20]. SPPs are electromagnetic surface waves at a dielectric–metal interface, coupled to the charge density oscillation in the metal surface [23]. SPPs offer the ability to focus light on nanoscales and are key elements in the development of subwavelength optical components with the added advantage of being compact and operating at much high frequencies [20].

In this paper, we design and optimize an electro-optic modulator using an electrically switchable VO2 double triple-cavity. We take advantage of the large refractive index contrast between $VO_2$ in the metallic and insulator phases. We design the nanorods to achieve the ENZ condition in the OFF state and thus a strong transmission. The silicon waveguide thus has a low loss in the OFF state while the waveguide experiences a very strong attenuation in the ON state resulting from both the high extinction coefficient of the $VO_2$ nanorods as well as the destructive interference of the nanorod cavity with VO2 in its metallic phase.

Using the finite element method, we investigate the optimal parameters for the nanorod cavity and obtain the geometry of nanorods which provide a strong modulation as well as a low insertion loss. We optimize our design such that in the ON state, the cavity destructively interferes with incoming signal resulting in a large modulation contrast. We thus present a compact (1.6μm x 1μm), low loss modulator with a modulation depth of 20.35 dB. (19.7 dB/μm) at the 1550nm telecommunications wavelength. We also propose a novel hexagonal arrangement of the nanorods to achieve a lower insertion loss while maintaining a strong modulation depth.

general device description

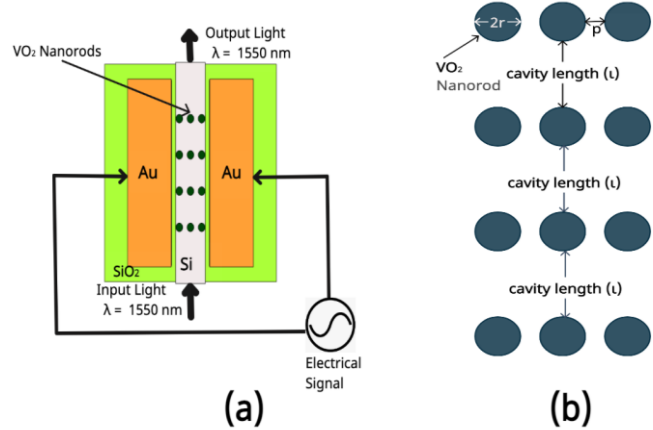

Fig. 1(a) Two-Dimensional schematics of the electro-optic modulator consisting of $VO_2$ nanorods embedded in a silicon strip waveguide(b) Zoomed in view of the arrangement of VO2 nanorod arrays showing 4 layers (3 cavities). The geometry can be tuned to achieve the ENZ condition in the OFF state.

Our nanorod modulator design consists of a Si strip waveguide with a thickness of 340nm and a width of 300nm with a nanorod cavity of $VO_2$ integrated in the middle of the waveguide. This waveguide supports a transverse magnetic-like (TM) mode. For the TM mode in our study, we use the effective medium theory to approximate the component of the permittivity tensor of the metamaterial formed using the $VO_2$ nanorods which is given by $\varepsilon_z = N\varepsilon_{VO_2} + (N-1)\varepsilon_{Si}$ where $\varepsilon_{VO_2}$ and $\varepsilon_{Si}$ represent the permittivity of VO2 and Si respectively, and $N = \pi(r/p)^2$ is the concentration of the $VO_2$ nanorods calculated in a uniform medium with a nanorod radius r and inter-distance p [15].

We take advantage of the huge refractive index change of $VO_2$ accompanying its first order Insulator to Metal transition to achieve optical modulation with the aid of an external electric field applied to the Gold electrodes as shown in Fig. 1a.

In this device, light of wavelength 1550nm is guided into the waveguides via a grating coupler (not shown on the figure). The light travels through the silicon waveguide unto the modulating section of the device with the $VO_2$ nanorod arrays integrated into the silicon waveguides.

The device operates in either an OFF state (without an electric field applied) or an ON state (with an electric field applied). In the OFF state, light travelling through the silicon waveguide interacts with $VO_2$ in the semiconductor phase. The cavity length and nanorod diameter are designed such that the metamaterial achieves the ENZ condition in the OFF state. This leads to a high transmission in the OFF state of the device and most of the light in the waveguide is transmitted from the input to the output port. In the ON state, $VO_2$ experience a phase change from the insulator to the metallic phase. With $VO_2$ in the metallic phase and an appropriate nanorod concentration, the effective permittivity approaches that of a metal making the double cavity more reflective and lossy.

The nanorod double cavity can tuned to further increase the modulation depth because in the metallic phase, the VO2 nanorods are highly reflective and act as hyperbolic mirrors. At a given cavity length, a standing wave builds up in the cavity leading to a maximum transmission resulting from a constructive interference typical of Fabry Perrot cavities. [17].

For our purposes, however, we tune the cavity to maximally interfere destructively with the incoming light signal to achieve a strong modulation index. The combined effect of high extinction coefficient of VO2 in the metallic phase and the destructive interference of the nanorod cavities lead to most of the light being blocked and thus significantly higher optical attenuation in the ON phase.

The modulation index is measure of how the modulated carrier signal differs from the unmodulated carrier signal and this can be obtained by measuring the difference in optical attenuation in the device in both states. The attenuation in dB is given by: Attenuation = $-10 \log_{10}\left(\frac{P_{output}}{P_{input}}\right)$ and the modulation depth is given by Attenuation (ON state) – Attenuation (OFF state). Using the finite element method, we investigate the optimal parameters for the nanorod cavity and obtain the geometric details of nanorods which provide a strong modulation as well as a low insertion loss. The final device parameters we obtain are nanorod radius (r = 38.9nm), cavity length (l = 250nm) and rod interdistance (g= 50nm). Our device is optimized for the C band of the telecommunications window.

Results and discussion

The design, simulation and geometric optimization of our nanorod modulation is carried using the Finite element method implemented in the commercial software COMSOL. The values of the refractive index of $VO_2$ used are obtained from variable angle spectroscopic ellipsometry as show in [6]. We use second-order scattering boundary conditions to terminate the computational domain and an extremely fine mesh size of 15nm is utilized within the nanorods which have the critical dimension in our device. We show using the finite element method that the effective medium theory which describes the average optical properties of metamaterials can be used to describe the behavior of a scarce $VO_2$ metamaterial nanorod array.

Fig. 2 shows a planar cross section of the electric field norm of the device in the OFF state. In this state, there is less attenuation compared to the ON state of the device which is characterized by greater attenuation after interaction with the nanorod triple-cavity.

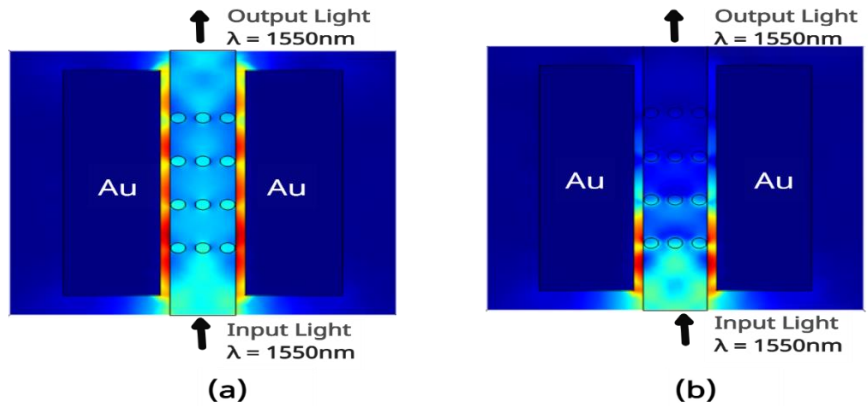

Fig.2 **Planar cross-section view of the modulator showing plot of the electric field norm.** (a) Device in the OFF state (b) Device in the ON state. High transmission is characteristic of the OFF state with the metamaterial meeting the ENZ condition and tuned for high transmission while in the ON state the metamaterial has a metal like permittivity and is tuned for minimum transmission.

A key metric in our design is the choice of the nanorod concentration (N). This parameter depends on the rod diameter and the rod inter-distance (p). For our study, we keep the rod inter-distance constant and perform a sweep on the rod diameter from 30nm to 80nm. We choose these values as we are limited by the width of the waveguide as well as current nanolithography dimensions.

The table below shows how the rod concentration and the component of the effective permittivity tensor of interest ($\varepsilon_z$) change with our rod diameter for our chosen value of p (50nm).

TABLE I

VARIATION OF EFFECTIVE PERMITTIVITY WITH NANOROD CONCENTRATION.

| Nanorod Concentration (N) | Diameter | Effective Permittivity, $\varepsilon_z$ (OFF State) | Effective Permittivity, $\varepsilon_z$ (ON State) |
| --- | --- | --- | --- |
| 0.1 | 17.841nm | 11.473 | 10.468 |
| 0.5 | 39.894nm | 8.927 | 3.902 |
| 0.73767246 | 77.812nm | 0.000 | -19.116 |
| 1 | 56.419nm | 5.743 | -4.307 |
| 1.90211983 | 77.812nm | 0.000 | -19.116 |

With a very small nanorod concentration (N), the effective permittivity of the metamaterial is close to that silicon and thus the average optical properties of the modulator are like those of a bare silicon nanowire. As the nanorod concentration increases the difference in the effective permittivity of the metamaterial in the ON and OFF phase grows and at a large enough value, the real part of the effective permittivity of the metamaterial is metal-like and thus highly reflective and lossy. For our purposes, we chose to tune the nanorod diameter such that in the ON phase the real part of the nanorod cavity approaches zero and in the OFF state, the real part is metal like. With this, the ENZ condition is met in the OFF state and not in the ON state as seen in Fig 3. In the ON state, the VO2 nanorod cavities are very lossy and very reflective and could be tuned to achieve a substantial loss compared to the ON state because they function more effectively as hyperbolic mirrors given the metal-like properties. Figure 3. shows how the real part of the effective permittivity of the metamaterial varies with the rod diameter in both states.

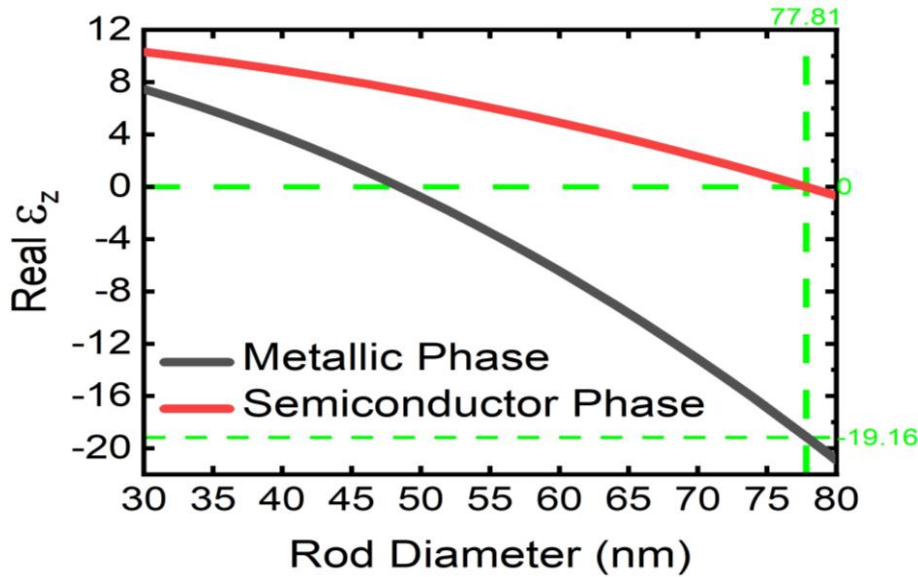

Fig. 3 Dependence of the Effective permittivity of the metamaterial on the nanorod diameter (inter-distance keep constant) with VO2 in both phases.

Given the key role played by the rod diameter and the cavity length (l) on the performance of our design, we swept the dimeter from 30nm to 80nm and the cavity length from 40nm to 400nm for incident light of wavelength 1550nm. We calculate the modulation depth in both the OFF (semiconductor) and ON (metallic) states of the device and plot these values on the same graph for evaluation.

From Fig.3a, the modulation depth increases as the nanorod diameter increases which could be explained by the increased difference in the effective permittivity of the metamaterial due to the higher nanorod concentration. It can also be observed that for a given nanorod diameter, changing the cavity length leads to significant changes in the modulation depth with the most pronounce change observed when the nanorod diameter is approximately close to 78nm and real part of the effective permittivity of the metamaterial approaches zero. From Fig2b, we can observe the tuning effect of the nanorod cavity with the modulation depth doubling as we tune the cavity length from 50nm to 250nm with the nanorod diameter kept constant at 77.81nm. This is because of the destructive interference of the nanorod cavities in the ON state. We leverage on this huge change in transmission to achieve a very high modulation depth in our device.

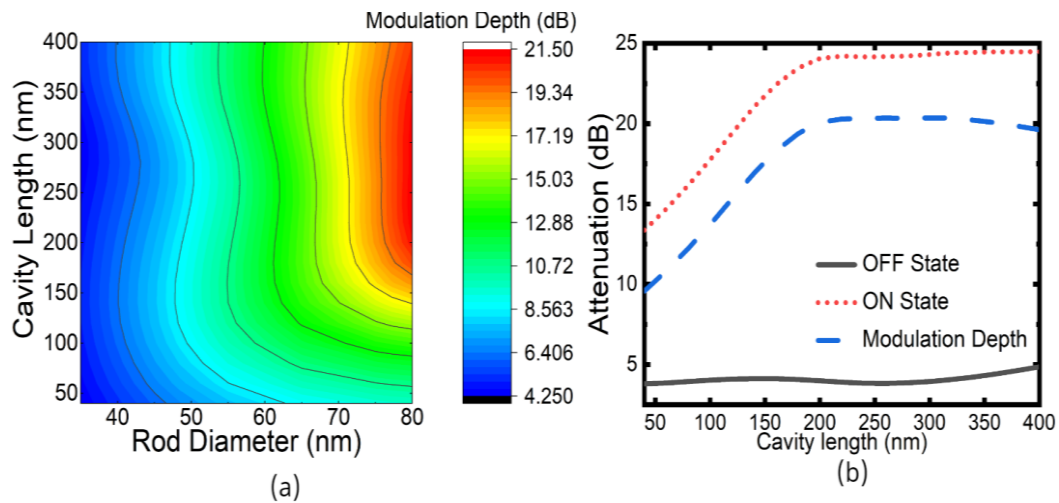

Figure 4 (a) Modulation depth of the cavity for different nanorod diameters and gaps. (b) Attenuation of device in ON state, OFF state and modulation depth for different cavity lengths

We also investigate the impact of the incident light wavelength on the insertion loss and modulation depth of our device by sweeping the incident wavelength from 1300nm to 1700nm. From Fig.5, the nanocavity modulator achieves a strong modulation depth across the 1300 to 1700nm wavelength range and achieves peak performance around the C band telecommunication window. This wide operating wavelength makes our device robust enough and non-susceptible to wavelength shifts resulting from ambient temperature variations in lasers.

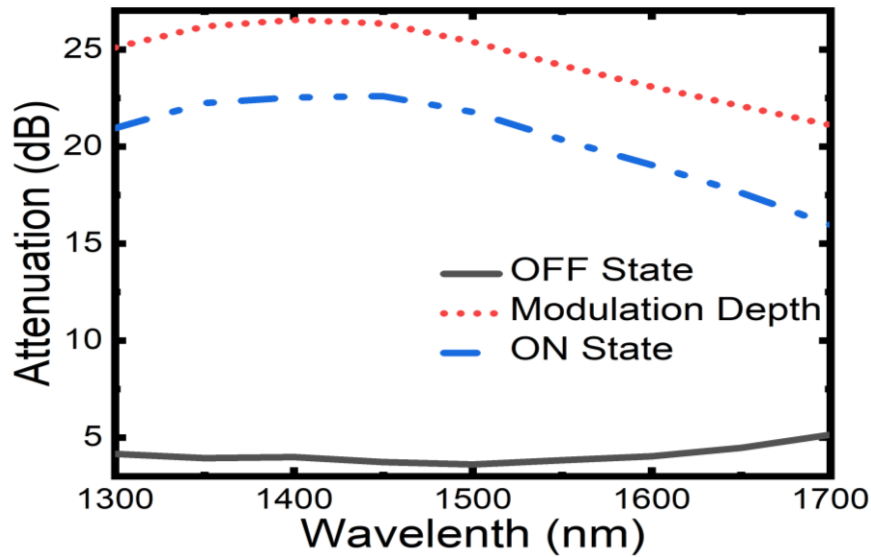

Fig. 5 Variation of attenuation of the modulator with different wavelengths of incident light (from 1300nm to 1700nm) in both the ON state and OFF state as well as the modulation depth of the device.

We further study the effect of adding each nanorod layer on the performance of our device. Using the finite element method, we quantitatively measure the modulation depth of device having one nanorod layer and subsequently add nanorod layers forming cavities of length 250nm.  study the effect of adding each nanorod layer to the overall performance. Fig.6a. shows there is a significant increase in modulation depth for each nanorod layer added. This comes at a cost however with the insertion loss of the device increasing are more layers are used. Fig.6 also shows how the electric field norm varies for each of the geometries in the ON state. It can be observed that by adding more nanorod layers, there is a greater attenuation because of the high extinction coefficient of $VO_2$ in its metallic phase. This study shows that we can adequately tune the performance of our modulator by introducing nanorod layers.

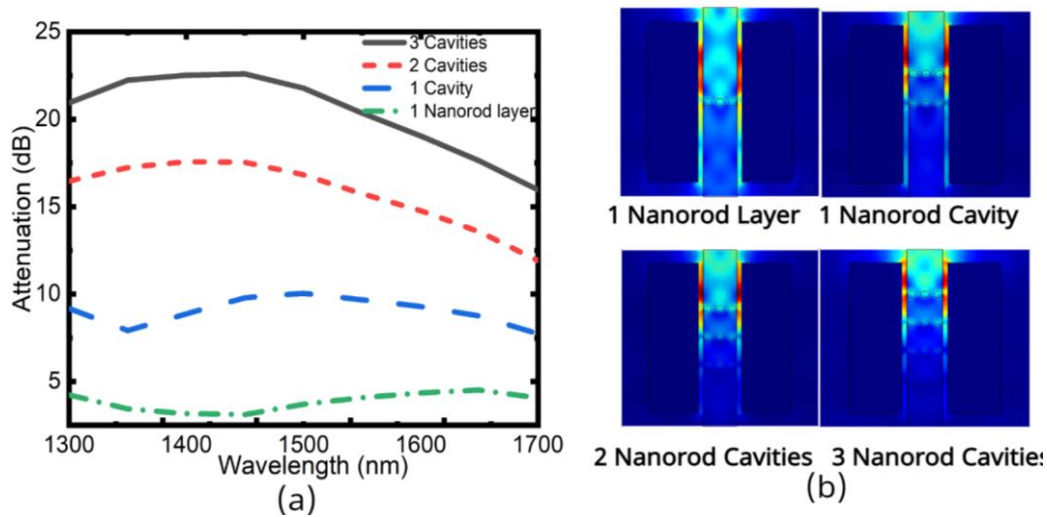

Fig. 6 (a) The Variation of the modulation depth with light of different wavelengths (from 1300nm to 1700nm) for 4 geometries each optained by adding a layer of VO$_2$ nanorods. (b) Planar cross section of 4 geometries showing electric field norm of each geometry in the ON state.

Finally, we propose a new geometry with a hexagonal array of nanorods as shown in figure 9b with the nanorods equidistantly positioned. This configuration helps us achieve a smaller insertion loss using a similar number of nanorods (13 nanorods compared to 12 nanorods in previous geometry). Fig. 7a shows the performance of our novel modulator as we perform a wavelength sweep from 1300nm to 1700nm. Figure 9a shows the insertion loss and modulation depth of the proposed design as well the difference in insertion loss of the hexagonal geometry compared to the triple cavity nanorod geometry. The proposed geometry shows a very low insertion loss and a good modulation depth and has a broad operation wavelength.

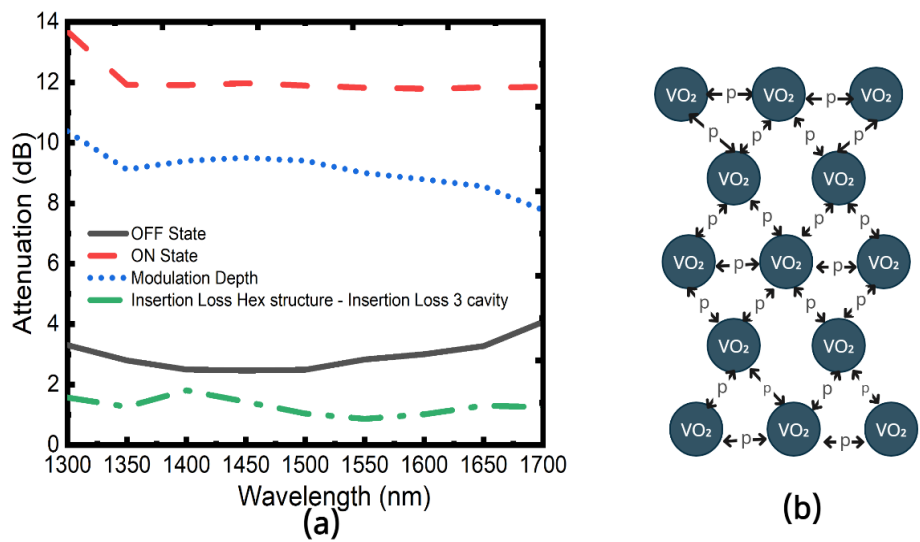

Fig. 7 (a) The Variation of the modulation depth with light of different wavelengths (from 1300nm to 1700nm) for Hexagonal nanorod array geometry as well as the difference in insertion loss between Hexagonal nanorod geometry and nanorod triple-cavity geometry. (b) Schematics of the arrangement of VO$_2$ nanorods in hexagonal array geometry.

Conclusion

In this paper, we have demonstrated a compact electro-optic modulator based on a VO$_2$ nanorod triple-cavity integrated inside a Silicon strip waveguide. The device has a footprint of 1.6umx 1um with the modulating section of dimensions 340 µm x 1 µm. The VO$_2$ nanorod cavities operate in the ENZ regime when the device is the OFF state and exhibits a strong non-linear response upon the application of an external electric field via the electrodes spotlighting how ENZ effects facilitate ultracompact and high speed modulator designs. We also carried out a study on the impact of each additional nanorod layer on the overall performance of the device giving us an insight on the contribution of each layer on the modulation depth. We then proposed a hexagonal geometry with equally spaced nanorods for a device with a significantly smaller insertion with a high modulation index and broad wavelength of operation than the triple-cavity geometry earlier introduced. Our work could be used in the design of high speed nanoscale modulators and ultrathin thermal switches.

References

[1] R Rajasekharan, TD Wilkinson, PJW Hands, Q Dai, Nanophotonic three-dimensional microscope, Nano letters 11 (7), 2770-2773 (2011)
[2] J.Yong, B. Hassan, Y. Liang, K. Ganesan, R. Ranjith, R. Evans, G. Egan, O. Kavehei, J. Li, G. Chana, B. Nasr & E. Skafidas, "A Silk Fibroin Bio-Transient Solution Processable Memristor," Scientific Reports, 7,14731 (2017)
[3] R Unnithan, Q Dai, TD Wilkinson, "Electro-optic characteristics of transparent nanophotonic device based on carbon nanotubes and liquid crystals," Applied optics 49 (11), 2099-2104, 2010
[4] Neira, A., Wurtz, G. and Zayats, A., 2018. All-optical switching in silicon photonic waveguides with an epsilon-near-zero resonant cavity [Invited]. Photonics Research, 6(5), p.B1.
[5] Markov, K. Appavoo, R. Haglund and S. Weiss, "Hybrid Si-VO_2-Au optical modulator based on near-field plasmonic coupling", Optics Express, vol. 23, no. 5, p. 6878, 2015.

[6] I. Sopko and G. Knyazev, "Optical modulator based on acousto-plasmonic coupling", Physics of Wave Phenomena, vol. 24, no. 2, pp. 124-128, 2016.
[6] A. D. Neira, G. A. Wurtz, P. Ginzburg, and A. V. Zayats, "Ultrafast all-optical modulation with hyperbolic metamaterial integrated in Si pho-tonic circuitry," Opt. Express 22, 10987–10994 (2014).27.
[7] A. Neira, N. Olivier, M. Nasir, W. Dickson, G. A. Wurtz, and A. V. Zayats, "Eliminating material constraints for nonlinearity with plasmonic metamaterials," Nat. Commun.6, 7757 (2015).
[8] C. L. Cortes and Z. Jacob, "Photonic analog of a van Hove singularity in metamaterials," Phys. Rev. B88, 045407 (2013)

[9] P. Sun, W. Shieh and R. Unnithan, "Design of Plasmonic Modulators with Vanadium Dioxide on Silicon-on-Insulator", IEEE Photonics Journal, vol. 9, no. 3, pp. 1-10, 2017.
[9] M. Talafi Noghani and M. Vadjed Samiei, "Propagation Characteristics of Multilayer Hybrid Insulator-Metal-Insulator and Metal-Insulator-Metal Plasmonic Waveguides", Advanced Electromagnetics, vol. 2, no. 3, p. 35, 2014. Available: 10.7716/aem.v2i3.222.
[10] B. Wu, A. Zimmers, H. Aubin, R. Ghosh, Y. Liu and R. Lopez, "Electric-field-driven phase transition in vanadium dioxide", Physical Review B, vol. 84, no. 24, 2011.
[11] Y. Salamin et al., "Author Correction: Microwave plasmonic mixer in a transparent fibre–wireless link", Nature Photonics, vol. 12, no. 12, pp. 790-790, 2018.
[12] W. Heni et al., "Nonlinearities of organic electro-optic materials in nanoscale slots and implications for the optimum modulator design", Optics Express, vol. 25, no. 3, p. 2627, 2017.
[13] X. Xu, C. Chung, Z. Pan, H. Yan and R. Chen, "Periodic waveguide structures for on-chip modulation and sensing", Japanese Journal of Applied Physics, vol. 57, no. 82, pp. 08PA04, 2018. Available: 10.7567/jjap.57.08pa04.
[14] A. Melikyan et al., "High-speed plasmonic phase modulators", Nature Photon., vol.8, no. 3, pp. 229-233, 2014.
[15] P. Markov, R. Marvel, H. Conley, K. Miller, R. Haglund and S. Weiss, "Optically Monitored Electrical Switching in VO2", ACS Photonics, vol. 2, no. 8, pp. 1175-1182, 2015.
[16] Y. Abate et al., "Control of plasmonic nanoantennas by reversible metal-insulator transition", Scientific Reports, vol. 5, no. 1, 2015.
[17] Mirsgafieyan, S.S., Guo, H. & Guo, J. Zeroth order fabry perot resonance enabled strong light absorption in thin silicon films of different metals and its application for color filters. IEEE photon. 8, 6804912 (2016)
[18] D. Gramotnev and S. Bozhevolnyi, "Plasmonics beyond the diffraction limit", Nature Photonics, vol. 4, no. 2, pp. 83-91, 2010.
[19] Y. Kim et al., "Phase Modulation with Electrically Tunable Vanadium Dioxide Phase-Change Metasurfaces", Nano Letters, vol. 19, no. 6, pp. 3961-3968, 2019. References
[20] Zhou, F. and Liang, C. (2019). Highly tunable and broadband graphene ring modulator. Journal of Nanophotonics, 13(01), p.1.
[21] S. Zouhdi, A. Sihvola and A. Vinogradov, Metamaterials and plasmonics. Dordrecht: Springer, 2009.
[22] S. K. Earl et al., "Tunable optical antennas enabled by the phase transition in vanadium dioxide", Opt.Exp., vol. 21, no. 22, pp.27503-27508, Dec. 2013.
[23] Cheng, D., 1992. Field And Wave Electromagnetics. 2nd ed. Reading, Mass.: Addison-Wesley.
[24] B.A. Kruger, A. Joushaghani, and J.K.S Poon, "Design of electrically driven hybrid vanadium dioxide (VO2) plasmonic switches" Opt. Exp., vol.20, no.21, pp.23598-23609, Oct. 2012.
[25] Q Dai, R Rajasekharan, H Butt, X Qiu, G Amaragtunga, TD Wilkinson, "Ultrasmall microlens array based on vertically aligned carbon nanofibers", Small 8 (16), 2501-2504 (2012)
[26] R Rajasekharan, H Butt, Q Dai, TD Wilkinson, GAJ Amaratunga "Can nanotubes make a lens array?" Advanced materials 24 (23), OP170-OP173 (2012)